\documentclass[%
 reprint,
 superscriptaddress,
 amsmath,amssymb,
 aps,
]{revtex4-2}

\usepackage{graphicx}% Include figure files
\usepackage{dcolumn}% Align table columns on decimal point
\usepackage{bm}% bold math
\usepackage[dvipsnames]{xcolor}
\begin{document}
%\preprint{APS/123-QED}

%\title{Absolute negative mobility in a one-dimensional overdamped system \\driven by active fluctuations}
\title{Active fluctuations drive inertia-free absolute negative mobility\\ in soft and living matter}
%\title{The simplest mechanism for going the wrong way:\\ absolute negative mobility in an overdamped system}
% Force line breaks with \\

\author{Karol Bia{\l}as}
\affiliation{Institute of Physics, University of Silesia, 41-500 Chorz{\'o}w, Poland}
\author{Peter H\"anggi}
\affiliation{Institute of Physics, University of Augsburg, D-86135 Augsburg, Germany}
\affiliation{Max Planck Institute for the Physics of Complex Systems, D-01187 Dresden, Germany}
\author{Jakub Spiechowicz}
\email{jakub.spiechowicz@us.edu.pl}
\affiliation{Institute of Physics, University of Silesia, 41-500 Chorz{\'o}w, Poland}

\begin{abstract}
Absolute negative mobility (ANM) is one of the most paradoxical transport phenomena in which a setup moves on average in a direction opposite to the applied force. The simplest system exhibiting this effect in minimal one-dimensional dynamics involves an inertial particle subjected to a constant bias while dwelling in a nonlinear, symmetric periodic potential in a nonequilibrium, nonstationary state generated by external driving. The necessity of inertia poses a significant conceptual barrier in predicting the ANM in a microscopic world of soft and living matter such as biological cells. In this work, we solve a two-decade old puzzle and demonstrate that none of the previously thought factors required for this effect are decisive. We show that the ANM may occur in a system composed of an overdamped particle in a piecewise linear symmetric periodic potential in an equilibrium state, provided that it is additionally driven by active fluctuations in the form of white Poisson shot noise. Our result may help to explain exotic transport behaviour emerging in biological cells where dynamics is typically overdamped and assisted by active fluctuations derived from various metabolic activities. It can also be exploited for effective separation strategies in a microscopic world, thus transforming fluctuations from a nuisance into a functional resource.

%Absolute negative mobility (ANM) is one of the most paradoxical transport phenomena in which a setup moves on average in a direction opposite to the applied force. According to the state of the art a simplest system exhibiting this effect in a minimal one-dimensional dynamics involves an \emph{inertial} particle subjected to a constant bias when dwelling in a \emph{nonlinear} symmetric periodic potential in a \emph{nonequilibrium} and \emph{nonstationary} state generated by an external driving. The necessity of inertia pose a significant conceptual barrier in predicting the ANM in a microscopic world of soft and living matter such as biological cells. In this work we solve a two-decade old puzzle and demonstrate that none of the previously though factors required for this effect are decisive. We show that the ANM may occur in a system composed of an \emph{overdamped} particle in a \emph{piecewise linear} symmetric periodic potential in an \emph{equilibrium} state provided that it is additionally driven by active fluctuations in the form of white Poisson shot noise. Our result may help to explain exotic transport behavior emerging in biological cells where dynamics is typically overdamped and assisted by active fluctuations derived from various metabolic activities. It can be also exploited for effective separation strategies in a microscopic world thus transforming fluctuations from a nuisance into a functional resource.

\end{abstract}

%absolute negative mobility, Brownian particle, active fluctuations
\maketitle

\section*{Introduction}

Transport phenomena in nonequilibrium systems sometimes defy our intuitive understanding of cause and effect. 
Absolute negative mobility stands out as perhaps the most paradoxical effect in which a setup persistently drifts opposite to an applied bias. It has been predicted in wide array of systems, such as particles moving through channels \cite{anm4,anm5,anm_added_Hanggi,anm21,anm27} or periodic potentials \cite{anm9,anm10,anm11,anm13,anm17,anm18,anm19,anm28,anm29,anm30,anm33,anm34,anm35,anm38,anm39,wisniewski}, Josephson junctions \cite{anm12}, tracers in laminar flows \cite{anm23,anm26,anm32,anm47}, chiral fluids \cite{kalz}, skyrmions \cite{shapiro} or random walk models \cite{anm27,anm6,benichou}. %\cite{anm4, anm5, anm_added_Hanggi, anm6, anm9, anm10, anm11, anm12, anm13, anm17, anm18, anm19, anm21, anm23, anm26, anm27, anm28, anm29, anm30, anm32, anm33, anm34, anm35, anm38, anm39, benichou, wisniewski, anm47}
%Absolute negative mobility \cite{anm4, anm5, anm_added_Hanggi, anm6, anm9, anm10, anm11, anm12, anm13, anm17, anm18, anm19, anm21, anm23, anm26, anm27, anm28, anm29, anm30, anm32, anm33, anm34, anm35, anm38, anm39, benichou, wisniewski, anm47} stands out as perhaps the most paradoxical effect in which a setup persistently drifts opposite to an applied bias. 
It challenges the foundations of linear response theory yet it has been successfully corroborated experimentally \cite{anc, anm40, anm41, anm42, anm43, barakov, anm44, anm45, anm46, prb}.
%Among these cases, absolute negative mobility (ANM) \todo{cite}, when a system persistently drifts apposite to an applied bias, stands out as perhaps the most paradoxical one which challenges the very foundations of linear response theory. Yet it has been successfully corroborated experimentally \todo{cite}. 
According to the state of the art a minimal system capable of exhibiting the ANM in a one-dimensional dynamics involves \emph{inertia}, a \emph{nonlinear} symmetric periodic potential and a \emph{nonequilibrium} and \emph{nonstationary} state generated by an external driving force \cite{anm9,anm10}. When it is subjected to a constant bias the interplay of these factors creates conditions under which the response of the system can be inverted.

These stringent requirements, in particular the necessity of inertia, pose a key constraint in predicting the ANM in a microscopic world of soft and living matter such as biological cells. It is dominated by viscous friction and consequently the corresponding dynamics is overdamped. Many unsuccessful attempts to discover a mechanism which could operate under such simplified setting, especially in a one dimension, have been made over the last two decades \cite{anm15, dimer, wrong_over, comment, anm24, chiny}. Progress has been made by using the phenomenological framework of discrete exclusion processes, however, ANM ceases to exist when probing a corresponding realistic continuous systems via the microscopic Langevin approach \cite{anm27}.  Moreover, in living cells there are hardly any systematic gradients or constant forces but rather active fluctuations originating from various metabolic activities \cite{nonthermal}. 
%driving molecular motors, cytoskeletal rearrangements and enzymatic reactions to namely only a few \todo{cite}. 
%Unlike thermal equilibrium ones which are Gaussian and characterized by fluctuation-dissipation theorem \todo{cite} active fluctuations are inherently nonequilibrium. 
Their study has emerged as one of the active frontiers in modern physics \cite{turlier, active4, active3, ezber, ariga, active1, sharma} since understanding how these hallmarks of biological systems alter transport properties is necessary for decoding the physical principles of life.

In this work we connect these two domains and propose a minimal model for the emergence of the ANM. We demonstrate that none of the previously thought ingredients necessary to generate this effect is essential. We show that the ANM can arise in a notably simple system consisting of an \emph{overdamped} particle in a one-dimensional \emph{piecewise linear} symmetric periodic potential in an \emph{equilibrium} state provided that it is additionally driven by active fluctuations in the form of white Poisson shot noise and explain the mechanism of this phenomenon. These fluctuations, intrinsically impulsive and non-Gaussian, mimic the intermittent forces encountered in living systems and active matter \cite{needle,gnesotto,turlier2}.

The significance of our result extends way beyond theoretical reductionism. It recast the ANM as a fundamentally broader and more accessible phenomenon being induced by active fluctuations ubiquitously generated by metabolic activities in living cells. In particular, the findings may provide a particularly simple physical mechanism for how molecular motors might reverse direction using active cellular fluctuations thus helping to decode exotic transport behavior in biology and suggesting that active noise is capable of driving intricate, bidirectional intracellular logistics.

Furthermore, the simplicity of our model facilitates direct experimental verification using existing setups, such as colloids in optical tweezers \cite{optical1,optical2,lattice} and Josephson junctions \cite{kautz,anm43}, to name only a few. These insights not only deepen our understanding of intracellular transport but also suggest novel strategy for engineering transport in microscale domain. They can be employed in highly selective separation devices that deliberately exploit this paradoxical response, transforming fluctuations from a nuisance into a functional resource \cite{anm28,anm46,hanggi2009}.

%Fluctuations appear in a wide array of physical systems. In addition to thermal fluctuations, which e.g lead to Brownian motion, in nonequilibrium environments there are also nonthermal fluctuations\cite{nonthermal1, nonthermal2, nonthermal3}. By definition, they are all fluctuations other than thermal ones. They may appear due to heating, stirring, or externally applied force, among other factors. In a living cell, there are two main sources of them, one of them are extra-thermal effects from spatially non-uniform exo- and endothermic reactions, which result in local heating and cooling, respectively\cite{nonthermal1}. The other is hydrodynamic in nature and arises from directional movement of motor proteins like actomyosin, dynein or kinesin and their cargo\cite{nonthermal1,carpet}.

\section*{Results}
\subsection*{Model}
We consider a system of an overdamped Brownian particle in a periodic potential $U(x)$ subjected to both active $\eta(t)$ and thermal $\xi(t)$ fluctuations. It can be modeled by the following dimensionless Langevin equation
\begin{equation}
    \dot{x} = -U'(x) + \eta(t) + \sqrt{2 D_T}\, \xi(t)
    \label{model}
\end{equation}
where the dot and prime denote differentiation with respect to time $t$ and position $x$, respectively. Thermal fluctuations $\xi(t)$ are modeled as $\delta$-correlated Gaussian white noise $\langle \xi(t)\xi(s) \rangle = \delta(t-s)$ of vanishing mean \mbox{$\langle \xi(t) \rangle = 0$}. $D_T$ represents thermal noise intensity satisfying the fluctuation-dissipation theorem \cite{kubo,marconi}. We refer the reader to Ref. \cite{praca_w_PRE} for all details of the scaling procedure. We assume that active $\eta(t)$ and thermal $\xi(t)$ fluctuations are uncorrelated, i.e. $\langle \xi(t)\eta(s) \rangle = \langle \xi(t) \rangle \langle \eta(s) \rangle = 0.$

\begin{figure}[t]
    \centering
    \includegraphics[width=0.9\linewidth]{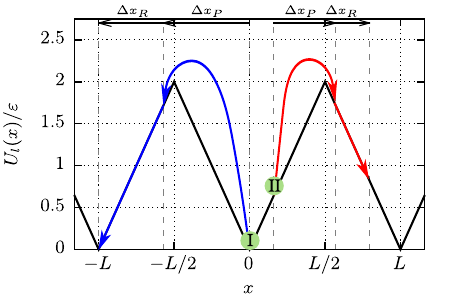}
    \caption{\textbf{Visualization of the particle dynamics.} Evolution of a Brownian particle dwelling in a piecewise linear potential $U_l(x)$ driven by a single $\delta$-spike of active fluctuations $\eta(t)$. $\Delta x_P$ stands for a distance covered by the particle due to the arrival of $\delta$-spike whereas $\Delta x_R$ represents the displacement because of the relaxation towards the potential $U_l(x)$ minimum. Two situations are exemplified: (i) the rare spiking regime $\lambda \to 0$, (ii) the case of finite $\lambda$.}
    %\caption{Toy model corresponding to one jump due to a single nonthermal fluctuation $\eta(t)$ in two variants: I, where the particle always starts at the minimum, and II, where the initial position is randomly distributed over the potential period.}
    \label{fig:0}
\end{figure}

In this study we consider two variants of the spatially periodic and symmetric potential $U(x) = U(x + L)$. The first form is a piecewise linear one given by 
\begin{equation}
	U_l(x) = 
    \begin{cases}
%      4\varepsilon x/L & x \in (0,\frac{L}{2}) \mod L, \\
%      %-2\varepsilon x/\pi+4\varepsilon &x/2\pi \in (\frac{1}{2},1) \mod 1
%       -4\varepsilon (x-L)/L &x \in (\frac{L}{2},L) \mod L.
		-4\varepsilon x/L, & x \in [-L/2,0) \mod L,\\
		4\varepsilon x/L, & x \in [0,L/2] \mod L.
    \end{cases}
    \label{U_l}
\end{equation}
It is visualized in Fig. \ref{fig:0}. The second one is the cosine potential 
\begin{equation}
    U_c(x)=-\varepsilon \cos(2\pi x/L).
    \label{U_s}
\end{equation}
Both variants have a barrier height of $\Delta U=2\varepsilon$ and the period $L$. Their minima are located at \mbox{$x = n L$}, $n \in \mathbb{Z}$. %The piecewise linear potential was chosen for its simplicity, as the goal of this letter was to present the minimal ANM model, and it is arguably the simplest spatially symmetric periodic potential. 

The process $\eta(t)$ originates from active character of the system, therefore we call it active fluctuations. It means they are inherently nonequilibrium. $\eta(t)$ is modeled in the form of a random sequence of $\delta$-shaped spikes with independent amplitudes $\{z_i\}$ distributed according to a common probability density $\rho(z)$, i.e. as Poisson white shot noise \cite{LuczkaBartussek,spiechowicz2014pre,bialas2020,mechanism,chaos,marathe}
\begin{equation}
    \eta(t)=\sum_{i=1}^{n(t)}z_i\delta(t-t_i),
\end{equation}
where $t_i$ are the arrival times of the Poisson counting process $n(t)$ \cite{feller1970}, 
%\cite{feller1970}
namely, the probability for occurrence of $k$ impulses in the time interval $[0,t]$ is given by
\begin{equation}
    Pr\{n(t)=k\}=\frac{(\lambda t)^k}{k!}e^{-\lambda t}.
\end{equation}
The parameter $\lambda$ determines how many $\delta$-pulses occur per unit of time on average and therefore it is called the mean spiking rate. 

We assume that the amplitudes $\{z_i\}$ are distributed according to the skew-normal density $\rho(z)$ \cite{azz,rijal2022,praca_w_PRE}, for the detailed form of this probability distribution cf. Eq. (\ref{skew_normal}) in the section Methods. This probability distribution is a generalization of the Gaussian probability density commonly encountered in physics and beyond due to the central limit theorem. This means that fluctuations can be bi-directional (both positive and negative) and that their variance $\sigma^2$ is independent of their mean $\zeta$. However, unlike normal distribution, it can be asymmetric around its mean and the degree of this skewness can be adjusted by independent parameter $\chi$. Such amplitude distribution can represent a wide range of processes such as e.g. a stochastic release of energy in a self-propulsion mechanism \cite{self1,self2,self3} or a random collision with a suspension of active microswimmers forming the active bath \cite{active4, active3, active1, active2} or a combination of both. The skew-normal distribution is completely characterized by its three statistical moments, namely, its average $\zeta$, variance $\sigma^2$ and skewness $\chi$.

The above active nonequilibrium fluctuations $\eta(t)$ have a finite statistical bias
\begin{equation}
	\langle \eta(t) \rangle = \lambda \langle z_i \rangle = \lambda \zeta
\end{equation}
% \begin{subequations}
% \begin{align}
% \begin{split}
% \langle\eta(t)\rangle&=\lambda \zeta, \\
% \end{split}\\
% \begin{split}
% \langle\eta(t)\eta(s)\rangle-\langle\eta(t)\rangle\langle\eta(s)\rangle&=\lambda(\sigma^2+\zeta^2)\delta(t-s), \\
% \end{split}
% \end{align}
% \label{eq_S_def}
% \end{subequations}
which is a product of the mean spiking rate $\lambda$ and the average amplitude $\zeta$. We note that for vanishing active noise $\eta(t) \equiv 0$ the system given by Eq. (\ref{model}) in the long time limit reaches a unique equilibrium state characterized by the Boltzmann-Gibbs probability density \cite{risken}.
%equation Eq. \ref{seq_2} does not have analytical solutions, as despite a piecewise periodic potential, the nonthermal noise amplitude distribution has an analytical but not closed form. In addition, the solutions can only be obtained in much simpler systems, e.g., with a harmonic trap \cite{exact} instead of a periodic potential. Therefore, we resort to precise numerical simulations using a Monte Carlo integration scheme on graphical processing units (GPUs) \cite{spiechowicz2015cpc}. A modified Euler-Maruyama scheme \cite{euler1,euler2} was employed. To prevent high mean spiking rate $\lambda$ causing the accumulation of numerical errors\cite{kim2007}, the thermal noise intensity was set to $D_T=0.01$ during the scaling process.
%\section{Results}

\subsection*{Absolute negative mobility in a one-dimensional overdamped system}
Unless stated otherwise in the remaining part of the work we stick to the following parameter regime \mbox{$\{ \varepsilon = 100, L = 2\pi, \zeta = 0.01, \sigma^2 = 20, \chi = 0.99, D_T = 0.01 \}$}.

\begin{figure}[t]
    \centering
    \includegraphics[width=0.9\linewidth]{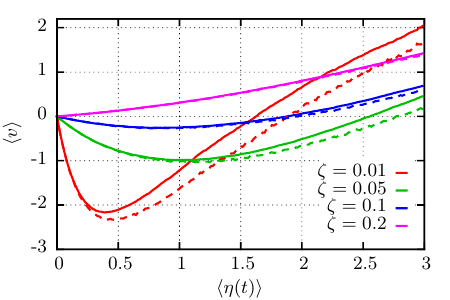}
    \caption{\textbf{Absolute negative mobility in the one-dimensional overdamped system.} The average velocity $\langle v \rangle$ of a Brownian particle as a function of the average bias $\langle \eta(t) \rangle = \lambda \zeta$ of active fluctuations depicted for different values of the mean amplitude $\zeta$. Other parameters read the barrier height $\varepsilon = 100$, thermal noise intensity $D_T=0.01$, variance $\sigma^2 = 20$ and skewness $\chi = 0.99$. The solid and dashed lines correspond to piecewise the linear $U_l(x)$ and cosine $U_c(x)$ potential, respectively.}
    \label{fig:1}
\end{figure}

In Fig. \ref{fig:1} we show the average velocity $\langle v \rangle$ of a Brownian particle dwelling in the piecewise linear $U_l(x)$ (solid lines) and cosine $U_c(x)$ (dashed lines) potential as a function of the statistical bias $\langle \eta(t) \rangle$ of active fluctuations for different values of the mean amplitude $\zeta$. We note that for the amplitude distribution $\rho(z)$ for which both positive $z > 0$ and negative $z < 0$ $\delta$-spikes are possible only the presented scaling $\langle \eta(t) \rangle = \lambda \zeta \to 0$ via the vanishing mean spiking rate $\lambda \to 0$ and $\zeta = \mbox{const.}$ implies $\eta(t) \to 0$ and therefore allows to study the response of the equilibrium system perturbed by active fluctuations $\eta(t)$. In contrast when $\langle \eta(t) \rangle \to 0$ via $\zeta \to 0$ with $\lambda = \mbox{const.}$ active fluctuations do not vanish $\eta(t) \neq 0$ and the system is inherently nonequilibrium.

The notable feature presented in Fig. \ref{fig:1} is the emergence of absolute negative mobility phenomenon \cite{anm9,anm10}, i.e. occurrence of the negative average velocity $\langle v \rangle < 0$ in the linear response regime $\langle \eta(t) \rangle \to 0^+$, despite the fact that (i) the statistical bias is positive $\langle \eta(t) \rangle > 0$, (ii) the mean amplitude of $\delta$-spikes is greater than zero $\zeta > 0$, (iii) the skewness of amplitude distribution $\rho(z)$ is positive $\chi > 0$ ($\rho(z)$ is tailed towards the positive values $z > 0$). It means that the particle travels on average in the direction opposite to the acting symmetry-breaking force $\eta(t)$. Moreover, since for sufficiently large bias $\langle \eta(t) \rangle \gg 1$ transport velocity must be eventually positive $\langle v \rangle > 0$ a distinct minimum is detected in the discussed characteristic when we observe the ANM.

The above finding must be contrasted with the state of the art \cite{anm9,anm10} which tells that the ANM can emerge in a one-dimensional system only if it is inertial, nonlinear and the symmetry breaking bias is applied to the setup in a nonequilibrium state guaranteed by e.g. an external driving force.  Our system is overdamped, piecewise linear and the biased active fluctuations $\eta(t)$ are applied to the setup given by Eq. (\ref{model}) initially in an equilibrium state. No additional forces are required. Thus, our work establishes active fluctuations as a novel mechanism for the ANM, overturning the previously assumed minimal requirements for its emergence, namely inertia and a pre-existing driven nonequilibrium state. For this reason our results contribute to our understanding of the ANM phenomenon and reveal another fascinating face of active fluctuations. 

Particularly, our findings supersede those in Ref. \cite{benichou} where the authors considered a two-dimensional exclusion process claiming that the ANM emerges due to the interactions with the environment of a tracer particle subjected to both, (i) a constant external force and (ii) an active force representing its propulsion mechanism. However, without the external bias their system is in nonequilibrium state induced by the active force. %The ANM appears only when the static force is additionally applied to such a driven system. 
The effect emerges due to the combined action of both the active and static force. This means that the active force does not give rise to the ANM by itself, but rather creates a necessary - yet not sufficient - condition for this effect. It rather corresponds to the original paradigm of the the ANM \cite{anm9,anm10}, where it was demonstrated that a nonequilibrium state of the system is required for the emergence of this paradoxical phenomenon. %Therefore, Ref. \cite{benichou} does not introduce any conceptual novelty in this respect.
%which without $\eta(t)$ reaches the equilibrium state.

%We start with the analysis of its dependence on the average force $\langle \eta(t)\rangle=\lambda \zeta$ under the condition that the amplitude distribution $\rho(z)$ is fixed, i.e. $\zeta=const$. In this scaling, only the mean spiking rate $\lambda$ changes. It can be understood as a change in the activity of the active fluctuations, e.g., for molecular motors it can model the change of concentration of ATP in the environment \cite{deguchi}.

\begin{figure}[t]
    \centering
    \includegraphics[width=0.9\linewidth]{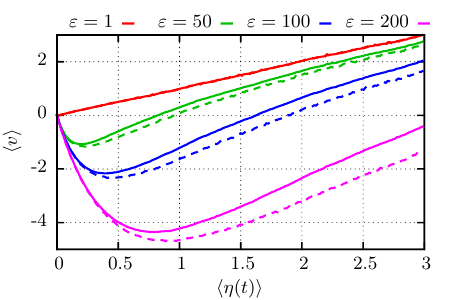}
    \caption{\textbf{Absolute negative mobility versus the potential barrier height.} The average velocity $\langle v\rangle$ as a function of average bias $\langle \eta(t)\rangle$ of active fluctuations for different values of barrier height $\varepsilon$. The mean amplitude is $\zeta=0.01$, variance $\sigma^2=20$, skewness $\chi=0.99$ and thermal noise intensity $D_T=0.01$. The solid and dashed lines correspond to the piecewise linear $U_l(x)$ and cosine $U_c(x)$ potential, respectively.}
    \label{fig:2}
\end{figure}

%\subsection{Toy model}
\subsection*{Mechanism of the effect}
In order to explain the mechanism behind the observed anomalous transport phenomenon let us consider how the arrival of a single $\delta$-spike with the amplitude $z$ affects the system. Since the role of thermal fluctuations in the studied parameter regime is minor we omit them. As it is illustrated in \mbox{Fig. \ref{fig:0}} the total particle displacement $\Delta x$ can be split into two components: the distance $\Delta x_P=z$ corresponding to the $\delta$-spike and the interval $\Delta x_R$ related to relaxation towards the nearest potential minimum. The latter is pronounced only when the particle has enough time to relax in the periodic potential before the next $\delta$-spikes arrives. Put differently, the mean time $\langle \tau_P \rangle = 1/\lambda$ between two consecutive $\delta$-spikes is significantly larger than the characteristic time $\langle \tau_R \rangle$ of the particle relaxation in the periodic potential \cite{praca_w_PRE}. In such a case the average velocity of the Brownian particle can be written as
\begin{equation}
    \langle v \rangle = \frac{\langle \Delta x \rangle}{\langle \tau_P \rangle} = \frac{\langle \Delta x_P\rangle +\langle\Delta x_R\rangle}{\langle \tau_P \rangle}= \lambda (\zeta + \langle \Delta x_R \rangle).
    \label{eq:av_v}
\end{equation}

From the above expression one can find the sufficient condition for the emergence of the ANM in the studied system. For the positive average amplitude $\zeta > 0$ the negative velocity yields 
\begin{equation}
	\langle v \rangle < 0 \Longrightarrow \langle \Delta x_R \rangle < -\zeta.
	\label{cond} 
\end{equation}
Since the particle displacement due to the relaxation in the periodic potential is bounded $\Delta x_R \in [-L/2,L/2]$ we conclude that the mean amplitude must obey $\zeta < L/2$. It is indeed the case in Fig. \ref{fig:1} when the ANM occurs. Moreover, the characteristic time $\langle \tau_R \rangle$ of the particle relaxation in the periodic potential depends on its steepness quantified by the barrier height $\varepsilon$. To satisfy the condition given by \mbox{Eq. (\ref{cond})} and permit the sufficiently large negative relaxation the potential $U(x)$ needs to have a high barrier $\varepsilon$. 
%so that $\langle \tau_P \rangle \ge \langle \tau_R \rangle$. 
It is confirmed in Fig. \ref{fig:2} where the average velocity $\langle v \rangle$ of the particle versus the mean bias $\langle \eta(t) \rangle$ is depicted for different barrier heights $\varepsilon$. We observe that the range of $\langle \eta(t) \rangle$ for which the negative mobility is detected increases for growing $\varepsilon$. Moreover, in such a case the minimal directed velocity $\langle v \rangle$ emerges for larger bias $\langle \eta(t) \rangle$ and its value decreases as well.
\begin{figure}[t]
    \centering
    \includegraphics[width=0.9\linewidth]{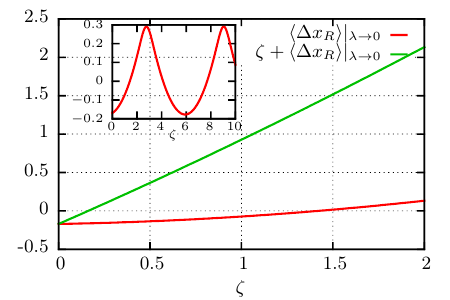}
    \caption{\textbf{The average relaxation and the mean displacement in the rare fluctuation regime.} The average relaxation distance $\langle \Delta x_R \rangle|_{\lambda \to 0}$ in the piecewise linear potential $U_l(x)$ for the limit of rare $\delta$-spikes (see Eq. (\ref{eq:x_R_0})) as a function of the mean amplitude $\zeta$ of active fluctuations $\eta(t)$. The variance reads $\sigma^2=20$, the skewness is $\chi=0.99$ and the barrier height $\varepsilon = 100$.}
    \label{fig:4}
\end{figure}

The average relaxation distance $\langle\Delta x_R\rangle$ can be calculated with the following formula
\begin{equation}
    \langle\Delta x_R\rangle = \int_{-L/2}^{L/2} dx \int_{-\infty}^{\infty} dz \int_0^{\infty} d\tau \phi(\tau) \rho(z) p(x) \Delta x_R(\tau,x+z).
    \label{eq:x_R_ran}
\end{equation}
where $\phi(\tau)= \theta(\tau) \lambda \exp(-\lambda \tau)$ is the probability density for random time interval $\tau$ between two consecutive $\delta$-spikes and $\theta(\tau)$ is a Heaviside step function. $p(x)$ stands for the distribution of the initial position of the particle. The displacement $\Delta x_R$ due to the relaxation in the piecewise linear potential given by Eq. (\ref{U_l}) can be written as 
\begin{equation}
    \Delta x_R(\tau, x)= -\text{sgn}(\mathbb{L}(x)) \min(4\tau\varepsilon/L,|\mathbb{L}(x)|),
    \label{eq:Delta_x_R}
\end{equation}
where $\text{sgn}$ is a sign function and the transformation $\mathbb{L}(x)=L\{(x+L/2)/L-\lfloor (x+L/2)/L\rfloor\}-L/2$ is used to fold the position after the $\delta$-spike to the interval $x \in [-L/2,L/2]$ by exploiting the periodicity of the potential. The $\lfloor \cdot \rfloor$ in $\mathbb{L}(x)$ represents the floor function, i.e. $\lfloor x \rfloor = \max(\{m \in \mathbb{Z}\,|\,m \leq x\})$. 

%$p(x)$ in Eq. (\ref{eq:x_R_ran}) is a probability distribution of initial position $x$ in the initial potential period. We consider two probability distributions, in the regime of rare impulses the assumption of initial position being at the potential minimum is valid, therefore $p(x)=\delta(x)$, where $\delta(x)$ is a Dirac delta. In this regime, integration over the initial position is simplified as $x=0$. The second considered probability distribution, $p(x)=P(x)$, requires knowledge of the real initial position distribution $p(x)=P(x)$ in the system. The easiest way to obtain it is by employing the Monte Carlo method, i.e., simulating a few $\delta$-impulses and truncating the particle position to the initial potential period.

In the linear response regime $\langle \eta(t) \rangle = \lambda \zeta \to 0$ via the rare $\delta$-spike limit $\lambda \to 0$ the expression (\ref{eq:Delta_x_R}) can be considerably simplified. In such a case the particle resides in the potential minimum $x = 0$ when the $\delta$-impulse arrives implying $p(x) = \delta(x)$ and the relaxation distance $\Delta x_R(z)$ no longer depends on the time interval $\tau$ due to the fact that $\tau$ is much longer than the characteristic time $\langle \tau_R \rangle$ of the particle relaxation. As a consequence,
\begin{equation}
    \langle \Delta x_R\rangle |_{\lambda\to0}=\int_{-\infty}^{\infty}dz \rho(z) \Delta x_R(z),
    \label{eq:x_R_0}
\end{equation}
and $\Delta x_R(z) = -\mathbb{L}(z)$ since then the particle always reaches the potential minimum during the relaxation. In Fig. \ref{fig:4} we show the above quantity as a function of the mean amplitude $\zeta$ of active fluctuations $\eta(t)$. We note that in the limit $\zeta \to 0$ the average relaxation distance $\langle \Delta x_R\rangle |_{\lambda\to0}$ is almost constant and therefore the critical amplitude $\zeta_c$ for which the ANM disappears reads $\zeta_c \approx -\langle \Delta x_R\rangle|_{\lambda\to0}(\zeta = 0) = 0.16$, see the green curve in Fig. \ref{fig:4}. In the inset we demonstrate that $\langle \Delta x_R\rangle |_{\lambda\to0}$ is the periodic function of $\zeta$ with the period equal to the spatial period $L$ of the potential and there are other regions where the average relaxation is negative, however, it is too small to overcome the larger mean amplitude $\zeta$ in Eq. (\ref{eq:av_v}) to generate the ANM.
\begin{figure}[t]
    \centering
    \includegraphics[width=0.9\linewidth]{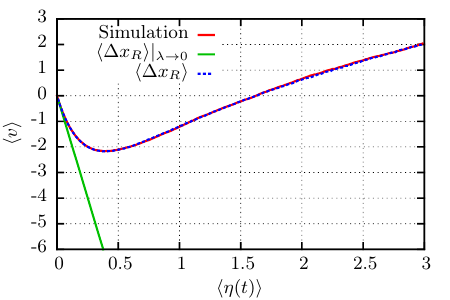}
    \caption{\textbf{The toy model versus the numerical simulations.} The average velocity $\langle v \rangle$ versus the mean bias \mbox{$\langle \eta(t)\rangle =\lambda \zeta$}. Comparison between the precise numerical simulations of the dynamics (\ref{model}) and the approach Eq. (\ref{eq:av_v}) with the average relaxation distance $\langle \Delta x_R \rangle$ given by  Eq. (\ref{eq:x_R_0}) and Eq. (\ref{eq:x_R_ran}). The average amplitude is $\zeta=0.01$, variance $\sigma^2=20$, skewness $\chi=0.99$, thermal noise intensity $D_T=0.01$ and the potential barrier height $\varepsilon=100$.}
    \label{fig:5}
\end{figure}

When moving away from the linear response regime $\langle \eta(t) \rangle = \lambda \zeta \to 0$, i.e. for growing frequency $\lambda$ of $\delta$-spikes, it is notable that the negative mobility effect is initially amplified despite the fact that the sampling of the amplitude distribution $\rho(z)$ tailed in the positive direction increases as well. 
In such a case the particle no longer resides in the potential minimum when the $\delta$-spike arrives and the relaxation distance $\Delta x_R(\tau,x+z)$ is influenced by the time $\tau$ between two consecutive $\delta$-impulses. To take these two factors into account one needs to use \mbox{Eq. (\ref{eq:x_R_ran})} instead of the simplified one given by formula Eq. (\ref{eq:x_R_0}) from which we find that in the studied parameter regime $\langle \Delta x_R\rangle|_{\lambda \to 0} < 0$. On the other hand, it is clear that in the limit $\lambda \to \infty$ the average relaxation distance vanishes $\langle \Delta x_R \rangle \to 0$ since the particle is constantly agitated by active fluctuations $\eta(t)$ and it does not have time to slide along the potential slope. From this observation we conclude that the term $\lambda \langle \Delta x_R \rangle$ in the average velocity $\langle v \rangle$ of the particle given by Eq. (\ref{eq:av_v}) is responsible for the emergence of the negative minimum. Note that it arises naturally from continuity of the function $\lambda \langle \Delta x_R \rangle$ which tends to zero in both limits $\lambda \to 0$ and $\lambda \to \infty$. %The maximal negative velocity is attained for the rate $\lambda_o$ obeying the relation 
%\begin{equation}
%\langle \Delta x_R \rangle(\lambda_o) + \lambda_o \frac{d\langle \Delta x_R(\lambda)\rangle}{d\lambda}\bigg|_{\lambda_o} = -\zeta.
%\end{equation}

In Fig. \ref{fig:5} we validate our approach to transport of the overdamped Brownian particle driven by active fluctuations. Its average velocity $\langle v \rangle$ calculated from precise numerical simulations of the underlying dynamics (\ref{model}) is compared there with the expression (\ref{eq:av_v}) with two variants of the average relaxation distance $\langle \Delta x_R \rangle$: (i) Eq. (\ref{eq:x_R_0}) in the linear response limit $\langle \Delta x_R\rangle|_{\lambda \to 0}$ and (ii) Eq. (\ref{eq:x_R_ran}). There is a perfect agreement between the latter and the simulation curve while the former correctly predicts the negative mobility in the linear response regime.

%As the thermal fluctuations were not included in the jump-relaxation model, we conclude that they are not required for the emergence of ANM. On the other hand, in the case of a continuous force described by the nogo theorem\cite{anm10}, the thermal fluctuations were necessary for ANM to appear. Therefore, the presented ANM mechanism breaks most of the requirements of the nogo theorem. Let us list the differences: the system may be overdamped instead of inertial; the environment may be piecewise linear instead of nonlinear; the thermal fluctuations are not required; without the active force $\eta(t)=0$, the system is at the thermodynamic equilibrium, while for the continuous force, it had to always be non-equilibrium.

\subsection*{Skewness and symmetry breaking}
\begin{figure}[t]
    \centering
    \includegraphics[width=0.9\linewidth]{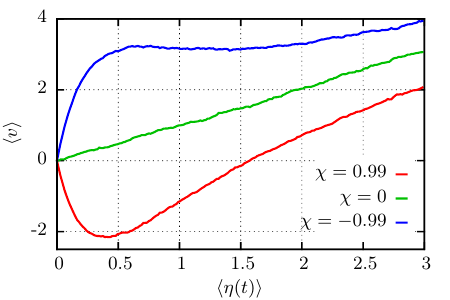}
    \caption{\textbf{The role of skewness $\chi$.} The average velocity $\langle v \rangle$ as a function of the mean bias $\langle \eta(t)\rangle=\lambda \zeta$ for different values of the skewness $\chi$ of the amplitude distribution $\rho(z)$. The average amplitude is $\zeta=0.01$, variance $\sigma^2=20$, skewness $\chi=0.99$, thermal noise intensity $D_T=0.01$ and the potential barrier $\varepsilon=100$}.
    \label{fig:9}
\end{figure}

\begin{figure}[t]
    \centering
    \includegraphics[width=0.9\linewidth]{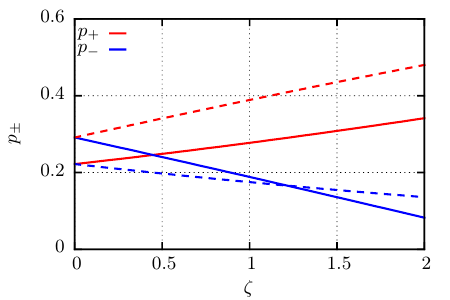}
    \caption{\textbf{Transfer probabilities over the nearest potential barriers.} The probability $p_+$ and $p_-$ to overcome the periodic potential $U(x)$ barrier in the positive and negative direction, respectively, due to the arrival of the $\delta$-spike as a function of the mean amplitude $\zeta$ in the rare spiking regime. Solid lines correspond to the skewness $\chi = 0.99$ whereas the dashed ones to $\chi = -0.99$. The variance is $\sigma^2=20$, while the potential barrier $\varepsilon = 100$.}
    \label{fig:7}
\end{figure}

Since in the discussed parameter regime the distribution $\rho(z)$ of active fluctuations $\delta$-spikes is asymmetric $\chi = 0.99$ one may naively think that the ANM emerges trivially solely due to the fact that the probability $p_-$ to overcome the potential barrier in the negative direction due to the arrival of $\delta$-pulse is greater than $p_+$ corresponding to the opposite one, namely
\begin{equation}
	p_- > p_+, \quad p_\pm = \pm \int_{\pm L/2}^{\pm \infty} dz\,\rho(z).
	\label{prob}
\end{equation}
However, the asymmetry of the amplitude distribution $\rho(z)$ is a necessary but not sufficient condition for the emergence of the ANM. It is evident from the inspection of Fig. \ref{fig:9} where we show the average velocity $\langle v \rangle$ of the particle versus the mean bias $\langle \eta(t) \rangle$ of active fluctuations for different values of the skewness $\chi$. In the studied regime the ANM ceases to exist for $\chi = 0$ when $\rho(z)$ is symmetric around its mean value $\zeta$. Likewise, it is not detected also for the negative skewness $\chi = -0.99$. It emerges only for $\chi = 0.99$ when the statistical symmetry of the amplitudes $z$ is broken in such a way that $\rho(z)$ is tailed into the positive direction, cf. Skew-normal distribution subsection in the Methods section.
%see Fig. \ref{fig:6}.
In Fig. \ref{fig:7} we show how the skewness $\chi$ of the amplitude distribution translates to the probability $p_+$ and $p_-$, c.f. Eq. (\ref{prob}), to overcome the periodic potential $U(x)$ barrier in the positive and negative direction, respectively, due to the arrival of the $\delta$-spike. We note that in the limit of small mean amplitude $\zeta \ll 1$, for which the ANM can emerge, $p_- > p_+$ only for $\chi = 0.99$ (solid lines) whereas for $\chi = -0.99$ (dashed lines) the opposite case $p_- < p_+$ takes place. However, $p_- > p_+$ is not a sufficient condition for the ANM to appear since it is satisfied up to $\zeta \approx 0.45$ whereas this effect no longer emerges for $\zeta > 0.16$, see Fig. \ref{fig:4}.
%In Appendix B we show that it is clearly not true. In particular, for the positive skewness $\chi = 0.99$, the condition $p_- > p_+$ is satisfied up to $\zeta \approx 0.45$ whereas the ANM ceases to exist already for $\zeta > 0.16$. However, while it is not a sufficient condition it certainly is a necessary one since the ANM does not emerge if $p_+ > p_-$, see e.g. the case of the negative skewness $\chi = -0.99$ in the studied regime.

\section*{Discussion}
In conclusion, we demonstrated emergence of the absolute negative mobility in a one-dimensional overdamped system consisting of a Brownian particle dwelling in a symmetric periodic potential  and driven by white Poisson shot noise. In doing so we revealed another fascinating face of active fluctuations. The origin of the detected anomalous transport phenomenon lies in the negative average displacement during the relaxation in the periodic potential. The necessary conditions for the operation of the discovered mechanism are (i) the average relaxation time significantly smaller than the characteristic time between two consecutive $\delta$-spikes of active fluctuations (ii) the asymmetry of the probabilities for the particle to overcome the potential barrier in the negative and positive directions. The former is satisfied for a high barrier whereas the latter is accomplished e.g. for a skewed amplitude distribution of $\delta$-spikes.

Our work enriches our understanding of the absolute negative mobility phenomenon by providing new minimal model for this effect. The presented results are vital from both fundamental and application point of view. They may help to explain exotic transport anomalies emerging in overdamped setups immersed in fluctuating environments such as biological cells or be exploited for effective separation of a diversity of nano and micro-sized particles of either natural or artificial origin. Our results may be experimentally verified in different setups ranging from colloids in optical tweezers \cite{optical1,optical2}, cold atoms in optical lattices \cite{lattice} to Josephson junctions \cite{kautz,anm43}, to name only a few.

\section*{Methods}
\subsection*{Skew-normal distribution}
\begin{figure}[t]
    \centering
    \includegraphics[width=0.9\linewidth]{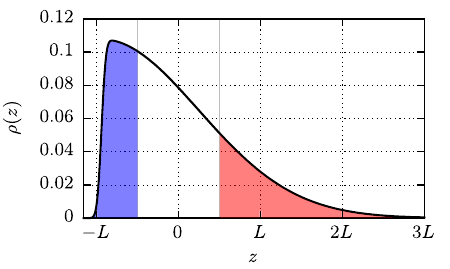}
    \caption{\textbf{Amplitude distribution $\rho(z)$.} Skew-normal distribution $\rho(z)$ of active fluctuations $\eta(t)$ amplitudes $z$ presented for mean $\zeta=0.01$, variance $\sigma^2=20$ and skewness $\chi=0.99$. Red and blue areas correspond to the probability of crossing the periodic potential $U(x)$ barrier in the positive and negative direction, respectively, due to the arrival of $\delta$-spike with the amplitude taken from $\rho(z)$. The spatial period reads $L = 2\pi$.}
    \label{fig:6}
\end{figure}

The probability density function of the skew-normal distribution reads \cite{azz}
\begin{equation}
    \rho(z) = \frac{2}{\sqrt{2\pi \omega^2}}e^{-\frac{(z-\mu)^2}{2\omega^2}} \int_{-\infty}^{\alpha[(z-\mu)/\omega]} \frac{1}{\sqrt{2\pi}}e^{-\frac{s^2}{2}} ds,
    \label{skew_normal}
\end{equation}
where the parameters $\mu$, $\omega$, and $\alpha$ represent position, scale and shape, respectively. They can be reformulated in terms of the statistical moments of the distribution $\rho(z)$, i.e., its mean $\zeta = \langle z_i \rangle$, variance $\sigma^2 = \langle (z_i-\zeta)^2 \rangle$ and skewness $\chi = \langle (z_i-\zeta)^3\rangle/\sigma^3$ \cite{sp_generacja,sp_generacja2}
\begin{subequations}
\begin{align}
\begin{split}
\alpha&=\frac{\delta}{\sqrt{1-\delta^2}}, \\
\end{split}\\
\begin{split}
\omega&=\sqrt{\frac{\sigma^2}{1- 2\delta^2/\pi}}, \\
\end{split}\\
\begin{split}
\mu&=\zeta-\delta\sqrt{\frac{2\sigma^2}{\pi(1-2\delta^2/\pi)}},\\
\end{split}
\end{align}
\label{eq_S_def}
\end{subequations}
where $\delta$ is defined as
\begin{equation}
    \delta=\text{sgn}(\chi)\sqrt{\frac{|\chi|^{2/3}}{(2/\pi)\{[(4-\pi)/2]^{2/3}+|\chi|^{2/3}\}}}.
    \label{eq_S_delta}
\end{equation}
In Fig. \ref{fig:6} we present the amplitude distribution $\rho(z)$ corresponding to the parameter regime studied in the work.

\subsection*{Numerical simulations}
Unfortunately, despite its apparent simplicity neither Eq. (\ref{model}) nor the corresponding Fokker-Planck-Kolmogorov-Feller one can be solved analytically even for a piecewise linear potential $U_l(x)$. It is due to the fact that active fluctuations amplitude distribution $\rho(z)$ has an analytical but not closed form. For this reason to analyze the studied system we resort to precise numerical simulations using a Monte Carlo integration scheme \cite{kim2007} implemented on a graphical processing unit (GPU) \cite{spiechowicz2015cpc}. The main quantity of interest is the long time average velocity of the particle
\begin{equation}
    \langle v \rangle = \lim_{t\to \infty} \frac{\langle x(t) \rangle}{t}. %- \langle x(0) \rangle}{t}.
\end{equation}

\subsection*{Funding}
This work has been supported by National Science Centre (NCN) Grant No. 2024/54/E/ST3/00257 (J.S.).

\subsection*{Data availability statement}
The data supporting findings of this study are available from the corresponding author upon request.

\subsection*{Code availability statement}
The code of simulations used in this study is available from the corresponding author upon request.

\subsection*{Author contributions}
J.S. contributed to the concept and design of the project. K.B. performed the research. K.B. and J.S. discussed the results. K.B, P.H. and J.S. wrote the manuscript.

\subsection*{Competing interest}
The authors declare no competing interests.
%\subsection*{End matter}
%\appendix

%\nocite{*}

%\section*{References}

\end{document}